\documentclass[prb,twocolumn,eqsecnum,superscriptaddress,showpacs,aps]{revtex4-1}
\usepackage{color}
\usepackage{amsmath,bm,verbatim}
\usepackage{graphicx,epsfig,braket,amssymb,amsfonts,tabularx,multirow,hyperref}
\usepackage[english]{babel}

\newcommand{\ba}{\begin{eqnarray}}
\newcommand{\ea}{\end{eqnarray}}
\newcommand{\be}{\begin{equation}}
\newcommand{\ee}{\end{equation}}
\newcommand{\bd}{\begin{displaymath}}
\newcommand{\ed}{\end{displaymath}}
\renewcommand{\v}[1]{{\bf #1}}
\newcommand{\bpm}{\begin{pmatrix}}
\newcommand{\epm}{\end{pmatrix}}
\newcommand{\nn}{\nonumber \\}

\graphicspath{{../figures/}{figures/}{../prx_submit/}}

\begin{document}
\title{Entanglement and corner Hamiltonian spectra of integrable open spin chains}
\author{Panjin Kim}
\affiliation{Department of Physics, Sungkyunkwan University, Suwon 16419, Korea}
\author{Hosho Katsura}
\affiliation{Department of Physics, Graduate School of Science,\\ The University of Tokyo, Hongo, Bunkyo-ku, Tokyo 113-0033, Japan}
\author{Nandini Trivedi}
\affiliation{Department of Physics, The Ohio State University, Columbus, Ohio 43210, USA}
\author{Jung Hoon Han}\email{hanjh@skku.edu}
\affiliation{Department of Physics, Sungkyunkwan University, Suwon 16419, Korea}
\begin{abstract}
We investigate the entanglement entropy (EE) and entanglement spectra (ES) of critical SU($N$) ($2\le N \le 4$) spin chains and other integrable models of finite length with the density matrix renormalization group method. For all models under investigation, we find a remarkable agreement of the level spacings and the degeneracy structure of the ES with the spectrum of the corner Hamiltonian (CS), defined as the generator of the associated corner transfer matrix. The correspondence holds between ES$^{(n)}$ at the $n$-th cut position from the edge of the spin model, and the spectrum CS$^{(n)}$ of the corner Hamiltonian of length $n$, for all values of $n$ that we have checked. The cut position dependence of the ES shows a period-$N$ oscillatory behavior for a given SU($N$) chain, reminiscent of the oscillatory part of the entanglement entropy observed in the past for the same models.
However, the oscillations of the ES do not die out in the bulk of the chain, in contrast to the asymptotically vanishing oscillation of the entanglement entropy. We present a heuristic argument based on Young tableaux construction that can explain the period-$N$ structure of the ES qualitatively.
\end{abstract}
\date{\today}
\maketitle

\section{Introduction}
\label{sec:intro}


Entanglement of quantum systems~\cite{EPR,Schrodinger} has been a flourishing theme in quantum optics and quantum information sciences over many years~\cite{zeilinger97,bennett00,milburn01,nielsen}. The entanglement idea has found place in the mainstream many-body physics in recent years, as a new potent tool for characterizing low-energy quantum states obeying the so-called area law of the entanglement entropy~\cite{srednicki93,trivedi94,wilczek94}.  Quite distinct from traditional analyses based on correlation functions, the entanglement approach offers new concepts such as the area law and the topological entanglement entropy~\cite{wen06,kitaev06} that define and classify the many-body states.

Entanglement-related ideas can be put to stringent tests in models of one-dimensional quantum magnetism. For instance, the logarithmic dependence of the entanglement entropy first derived using the conformal field theory~\cite{wilczek94} was later verified in the numerical investigation of one-dimensional lattice spin models at criticality~\cite{vidal03}. Later refinement of the entanglement entropy calculation by Calabrese and Cardy of a conformal system with boundaries~\cite{calabrese04} was also confirmed by numerical calculations on various spin chain models. As a nice by-product, central charges of the conformal field theory governing the low-energy dynamics of a given spin chain problem can be extracted accurately from fitting the numerical data to the Calabrese-Cardy (CC) formula~\cite{solyom07,solyom08,calabrese10,xavier11,xavier12,illuminati13,kaul15}.

Entanglement entropy is related to the entanglement spectrum~\cite{haldane08} in the same way that the energy spectrum governs the thermodynamic entropy in quantum statistsical mechanics. Given a particular bi-partition of the quantum system into two, the reduced density matrix of the subsystem can be expressed as the exponential of some Hermitian operator, dubbed the entanglement Hamiltonian~\cite{haldane08}. The entanglement Hamiltonian endows one with a full knowledge of the entanglement structure in the quantum system. It is not always possible, however, to have the explicit form of the entanglement Hamiltonian even though the physical Hamiltonian itself is well-known. Through detailed density matrix renormalization group (DMRG) calculations and exact diagonalization studies, we suggest that the form of the entanglement Hamiltonian is given by the generator of the corner transfer matrix~\cite{baxter} which is so called corner Hamiltonian, for a wide range of integrable spin chain models, summarized as a statement:
\\

{\it For integrable one-dimensional models with the unique ground state, whether gapped or critical, the low-lying levels of the entanglement Hamiltonian for an arbitrary cut position is given by those of the corresponding corner Hamiltonian}.
\\

\noindent Our conclusion differs from several related field-theoretical analyses~\cite{cho16,cardy16} in that the latter approaches do not yield such site-specific constraints on the entanglement Hamiltonian. It also differs from earlier works that focused only on the bipartition cut made half-way on the chain and assumed non-critical models~\cite{peschel98}. Our work generalizes these earlier claims to site-by-site correspondence for each $n$, valid for integrable models whether gapped or not.

The other important set of findings we provide in this paper concern the entanglement spectra in SU($N$) spin chains. With advances in cold atom technology, the SU($N$) spin model with $N$ greater than 2 are fast becoming realistic test beds for novel ideas in quantum magnetism. In cold atomic gases usually the focus has been on the two hyperfine states of $^6$Li~\cite{hulet15} and $^{40}$K \cite{esslinger13,zwierlein16} in order to emulate the behavior of electrons with $S=1/2$ in materials. However, both in materials because of multi-orbitals and in atomic gases with larger nuclear spin it is possible to encounter much richer phenomena. For example, the fermionic strontium isotope $^{87}$Sr has a nuclear spin $I=9/2$ and $^{173}$Yb has $I=5/2$ that make it possible to realize SU($N\!=\!2I\!+\!1$) different from the familiar SU(2) for electrons~\cite{rey14,takahashi12}.

The entanglement entropy of SU($N$) Heisenberg spin chains whose central charge is $c_N = N-1$ has been investigated earlier~\cite{greiter08}. Quite surprisingly, the site dependence of the entanglement spectrum has never been subject to the same level of scrutiny.
Here we report several findings of the site-by-site entanglement spectrum in SU(2) through SU(4) spin chain models of finite length. A characteristic oscillation in $n$ mod $N$ for the entanglement spectra is found, and argued to be the origin of the period-$N$ oscillation in the entanglement entropy noted earlier~\cite{solyom07,solyom08,kaul15}.

We show that the degeneracy in the low-lying entanglement levels can be understood from group-theoretical considerations as if they are the physical energy levels of a finite quantum system. These findings of the entanglement spectrum oscillations are not yet understood within the exact solution approach of the SU($N$) spin chain problem. It thus appears that mystery of the oscillations of the entanglement spectra
of SU($N$) spin chains still remains despite decades of intense research. We present here a heuristic argument based on Young tableaux construction that can explain the period-$N$ structure of the ES qualitatively.

The paper is organized in the following way. In Sec.\;\ref{sec:background} the basic concept of corner transfer matrix and reduced density matrix in conjuction with DMRG are reviewed briefly. In Secs. \ref{sec:SU(2)} and \ref{sec:SU(3),SU(4)} we summarize the DMRG findings on the ES of the critical SU($N$) chains for $N=2$ (Heisenberg $S=1/2$ spin chain), $N=3$ (Uimin-Lai-Sutherland model~\cite{uimin,lai,sutherland}), and $N=4$ (Kugel-Khomskii model~\cite{KK,KK-su(4)}). Periodic features in the site-by-site ES for each SU($N$) chain, and the equivalence of the ES to the corresponding corner Hamiltonian spectrum, are emphasized. Section \ref{sec:CH-ES} highlights the equivalence of the entanglement and the corner spectra for other integrable models.
We summarize and discuss possible future directions in Sec. \ref{sec:summary}.

\section{Reduced density matrix and corner transfer matrix}
\label{sec:background}

This section is devoted to a review of several concepts that will form the center of discussion in subsequent section, and how DMRG serves as a powerful method to extract such information for quantum spin models.

For a pure state $|\psi \rangle$, say a ground state of a spin chain model of length $L$, one can find its Schmidt decomposition~\cite{nielsen}

\begin{align}\label{eq:schmidt}
  |\psi \rangle = \sum\limits_\alpha \lambda_\alpha |u_{A,\alpha} \rangle \otimes |v_{B,\alpha} \rangle .
\end{align}
Here $|u_{A,\alpha}\rangle$ and $|v_{B,\alpha}\rangle$ are called left and right Schmidt vectors, and written entirely in terms of variables belonging to subsystems $A$ and $B$, respectively. The non-overlapping regions $A \cup B$ comprise the whole system. Due to a theorem of matrix algebra, one can be assured that $\lambda_\alpha$ is a non-negative real coefficient for each index $\alpha$, which runs over the smaller of the number of degrees of freedom in $A$ and $B$.

In a quantum chain of length $L$ with open ends, the line separating the two subsystems can be drawn along an arbitrary bond position $n$ ranging from 1 to $L-1$. The Schmidt eigenvalues $\lambda_\alpha$ depend on the choice of the bipartition,
as explicitly written

\ba \label{eq:pos-schmidt} |\psi \rangle = \sum\limits_\alpha \lambda_{n,\alpha} |u_{n, \alpha} \rangle \otimes |v_{L-n, \alpha} \rangle . \ea
The index $\alpha$ has the range that equals the smaller of the two subsystem dimensions for a given partition $n$.
Most of the analyses in subsequent sections will be concerned with site-selective Schmidt eigenvalues $\lambda_{n, \alpha}$ of integrable spin chains. Reduced density matrix (RDM) for the subsystem of length $n$ and the von Neumann entropy, also known as the entanglement entropy, are subsequently defined from the Schmidt decomposition of $|\psi\rangle$ as

\ba \label{eq:RDM}
  \rho^{(n)} &=& \sum\limits_{\alpha'} \langle v_{L\!-\!n,\alpha'} |\psi \rangle \langle \psi| v_{L\!-\!n,\alpha'} \rangle
         = \sum\limits_\alpha \lambda_{n,\alpha}^2 |u_{n,\alpha} \rangle \langle u_{n,\alpha} |, \nn
           S^{(n)} &=& -\text{Tr} \left[ \rho^{(n)} \log \rho^{(n)} \right] = - \sum_\alpha \lambda_{n,\alpha}^2 \ln \lambda_{n,\alpha}^2 .
\ea

As with the thermodynamic entropy, the entanglement entropy encodes an ``average" aspect of the system under investigation,
whereas a more ``microscopic" information can be found by looking at the individual Schmidt eigenvalues $\lambda_{n,\alpha}$. The entanglement Hamiltonian $H^{(n)}_{\text{ent}}$ for the subsystem of length $n$ can be defined in terms of RDM via

\ba \rho^{(n)} \propto e^{- H_{\text{ent}}^{(n)} } . \ea
Eigenvalues of the entanglement Hamiltonian given by

\ba \label{eq:es} ({\rm ES})^{(n)}_{\alpha}  =  -\ln \lambda_{n,\alpha}^2 , \ea
form what is known as the entanglement spectrum (ES)\;\cite{haldane08}.
Note that the entanglement entropy is not a linear function of ES, but assumes a slightly awkward relation $({\rm EE})^{(n)} = \sum_\alpha e^{-{\rm (ES)}^{(n)}_{\alpha} } ({\rm ES})^{(n)}_{\alpha}$. For this reason we introduce another quantity that is more directly related to the EE, which we call weighted entanglement spectrum (WES) and is given by the set of values

\ba \label{eq:wes}
({\rm WES})^{(n)}_{\alpha} &=& -\lambda_{n, \alpha}^2 \ln \lambda_{n, \alpha}^2 . \label{eq:WES}
\ea
ES and WES are the two main quantities of analyses in the following sections.
DMRG is an ideal tool for investigating the quantities defined in Eq.~(\ref{eq:RDM}) through Eq.~(\ref{eq:wes}).
To understand how DMRG works, assume we are to approximate the ground state $| \psi_\text{G} \rangle $ of a certain Hamiltonian, by a variational state

\begin{align}\label{eq:approx_gs}
  | \bar{\psi}_\text{G} \rangle = \sum\limits_{\alpha=1}^{m \leq d_A} \sum\limits_{j=1}^{d_B} a_{\alpha,j} |U_{A,\alpha} \rangle \otimes |V_{B,j} \rangle,
\end{align}
where $d_A$ ($d_B$) is the dimension of the sub-Hilbert space on the partition $A$ ($B$) and $|U_{A,\alpha} \rangle$ ($| V_{B,j} \rangle $) forms an orthonormal basis of dimension $d_A$ ($d_B$) in $A$ ($B$). The goal is to find $| \bar{\psi}_\text{G} \rangle$ that minimizes the 2-norm distance $\left|\left| |\psi_\text{G} \rangle - | \bar{\psi}_\text{G} \rangle \right|\right|$ by varying $a_{\alpha,j}$ and $|U_{A,\alpha} \rangle$.
The solution found by White~\cite{dmrg} is that $| \bar{\psi}_\text{G} \rangle$ expanded in the Schmidt basis (left and right Schmidt vectors) weighted by $m$ largest Schmidt eigenvalues $\lambda_\alpha$,

\begin{align}\label{eq:approx_schmidt}
  | \bar{\psi}_\text{G} \rangle = \!\!\!\!\!\!\! \sum_\alpha^{m\leq \text{min}(d_A,d_B)} \!\!\!\!\!\!\! \lambda_\alpha |u_{A,\alpha} \rangle \otimes |v_{B,\alpha} \rangle ,
\end{align}
is what best approximates the true ground state for any given $m$.
An important point in connection with quantum entanglement is that in the course of a simulation, the DMRG algorithm guarantees to find the Schmidt decomposition of the approximated ground state for arbitrary bipartition~\cite{schollwock11}.
One therefore simply needs to read off $\lambda_{\alpha}$ at each cut position $n$, which successively gives the knowledge of EE$^{(n)}$, ES$^{(n)}$, and WES$^{(n)}$.


Let us now turn our attention to the corner Hamiltonian (CH) and the corner transfer matrix (CTM).
It has been known that the RDM for a half-infinite chain is equivalent to the fourth power of the CTM (up to a constant)\cite{nishino97,nishino97book}.
An important observation dating back to the pre-DMRG era is that CTM for the integrable model can be related to the corner Hamiltonian through $C=e^{-\kappa' K}$~\cite{baxter81,baxter}\footnote{Baxter showed CTM could be exponentiated, but the term ``corner Hamiltonian'' has not been introduced at that time.}. Derivation of $K$ for several two-dimensional classical models are available (see e.g. Ref. \onlinecite{davies88}). Since in the infinite-size limit $\rho^{(n)} = e^{-H_{\rm ent}^{(n)} } \propto C_n^4$ and $C_n \propto e^{-\kappa' K^{(n)}}$, we arrive at the identity

\ba H_{\rm ent}^{(n)} = \kappa K^{(n)} ,  \label{eq:CH-ES_connection}
\ea
with the identification $\kappa'=4 \kappa$.
This connection has been emphasized and numerically proven explicitly by Peschel \emph{et al.} in the transverse Ising model and the XXZ Heisenberg model away from critical points, and for the particular case of bipartition $n=L/2$~\cite{peschel98}.

Several key questions remained unexplored, though. First of all, the proof of the equivalence between the ES and the CS becomes dubious at the critical point because the eigenstates of the CTM are un-normalizable and not well-defined except for the models reducible to free fermions~\cite{peschel04,peschel09}. Secondly, it was not understood if the ES$=$CS correspondence holds for arbitrary bipartition $n$.
In this paper we confront both these issues head-on using the DMRG.  In the next two sections, direct comparison of the ES and the CS at any finite segment of length $n$ (length of the subsystem $A$) in SU($N$) critical spin chains will be numerically examined.

\section{SU(2) critical spin chain}
\label{sec:SU(2)}

\subsection{Numerical results}

\begin{figure*}[htbp]
  \centering
    \includegraphics[width=0.90\textwidth]{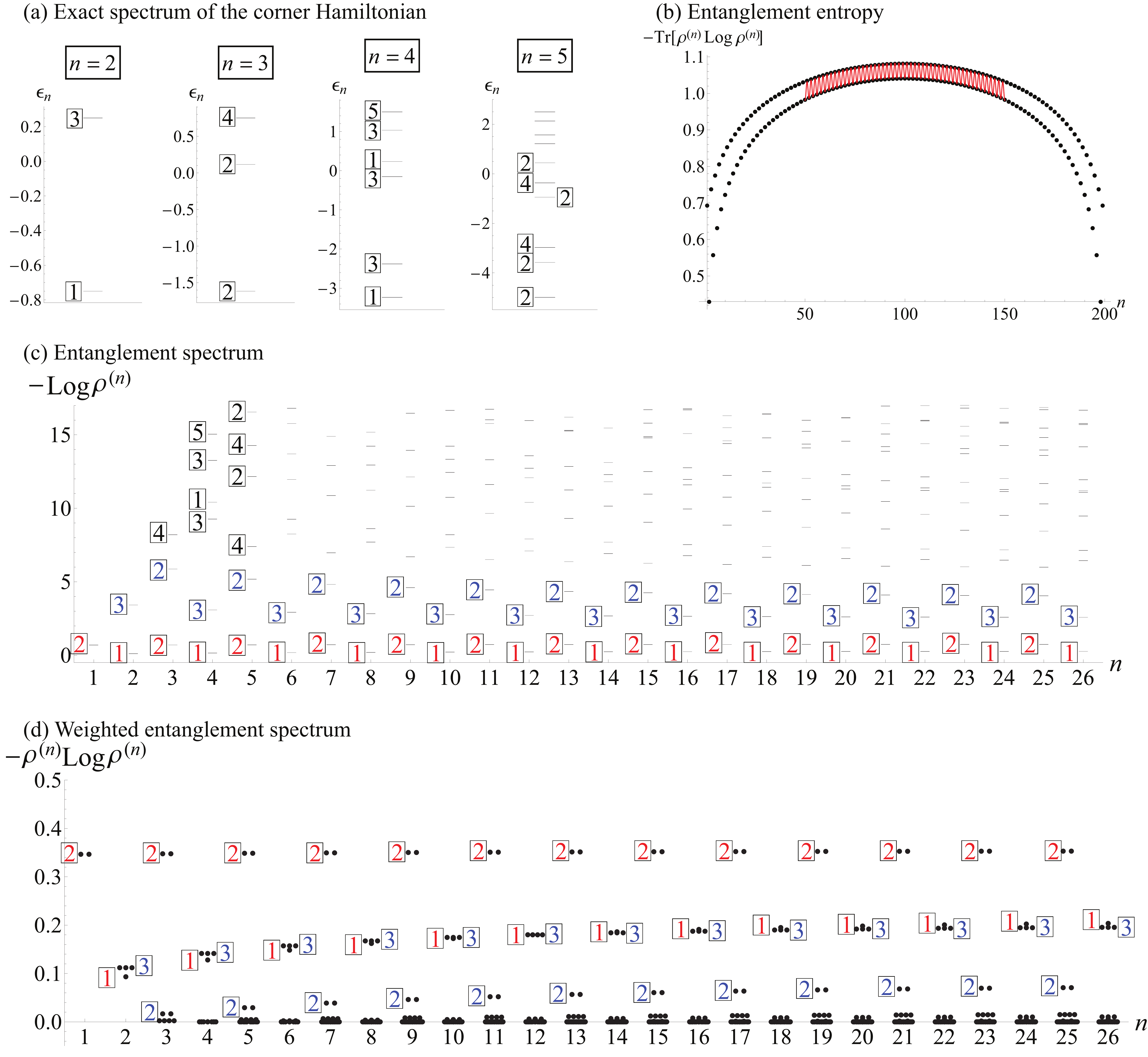}
   \caption{(color online) SU(2) critical spin chain. (a) Spectrum of SU(2) CH of length $n$.
   (b) EE, (c) ES, and (d) weighted ES at the $n$-th cut of the ground state for a spin chain of length 200.
   Degeneracy of the levels in (a), (c), (d) is indicated inside the box next to the horizontal bar. The red solid fit in (b) is obtained from  Eq. (\ref{eq:1+1cft}) with $c_N=1$ and $\Delta_1=1/2$.
   The ES alternate periodically between even and odd sites in the bulk as one can see from similar ES at $n=23$ and $n=25$ (odd $n$), and at $n=24$ and $n=26$ (even $n$). Entanglement levels with higher values of WES dominate the EE. For odd $n$, the first two levels, each doubly degenerate, dominate the EE. At even $n$, it is the first singlet and the second triplet levels that dominate. Other entanglement levels have WES values that accumulate around zero as shown in (d).}
\label{fig:su2_ed_dmrg}
\end{figure*}

We examine properties of the SU(2) critical spin chain written in the fundamental representation of the SU(2) algebra ${\bm S}=(1/2) \,\bm \sigma$ ($\bm \sigma$=Pauli matrix),

\begin{align}
  &H_{\rm SU(2)}^{(L)} = \sum\limits_{i=1}^{L-1} h_{\text{SU}(2),i}, \nn
  &h_{\text{SU}(2),i} = \bm \sigma_i \cdot \bm \sigma_{i+1}.
\label{eq:SU(2)-model}
\end{align}
Open boundary condition (OBC) for a chain of length $L$ is assumed. Ground states of the chain for lengths $L =60, 200$ are worked out by DMRG with the maximum number of Schmidt states kept being 1500.
ITensor library~\cite{itensor} is used in realizing the single-site DMRG with the noise algorithm~\cite{white05}.
Calculations of the ES for the $L=60$ chain was indistinguishable from the $L=200$ results, which gives us the confidence that the bond dependence of the ES reported in Fig. \ref{fig:su2_ed_dmrg}(c) should hold for arbitrary length of the SU(2) chain with open boundaries. One can easily obtain the EE and ES for every cut starting from one edge after the completion of the DMRG iteration.  Figure \ref{fig:su2_ed_dmrg}(c) shows the ES$^{(n)}$ of $L=200$ chain obtained by DMRG at the $n$-th bond, between sites $n$ and $n+1$, leaving the subsystem consisting of $n$ spins. For instance $n=1$ means only a single spin at the far left edge has been isolated as the subsystem. The two (up and down) spin degrees of freedom at the $n=1$ site naturally explains the double degeneracy of the ES. For the $n=2$ cut we see one single, and one triply degenerate ES levels.

By comparison we plot the spectrum of the corner Hamiltonian, defined as

\begin{align}
K_{\rm SU(2)}^{(n)} = \sum_{i=1}^{n-1} i \, h_{\text{SU}(2),i},
\label{eq:SU(2)-CH}
\end{align}
for a chain of length $n$.
In Fig. \ref{fig:su2_ed_dmrg}(a), the corner Hamiltonian spectrum is plotted for several $n$'s.
It is immediately clear that the corner spectrum for $n=2$ have the same singlet-triplet degeneracy structure found in the ES at the same $n$. In fact, inspection of Fig. \ref{fig:su2_ed_dmrg}(a) and (c) reveals that such correspondence of ES and spectrum of the CH extends without fault all the way up to $n=5$. (We checked the exact correspondence all the way up to $n=10$.)

Of course the degeneracy number is the dimension of all possible irreducible representations of the SU(2) and that the same degeneracy numbers appear in both the spectrum of entanglement and CH are not that surprising since the global SU(2) must be a symmetry of both the wave function of the CH and the left Schmidt eigenstate. The nontrivial aspect lies, in our opinion, rather with the {\it identical hierarchy of the levels} in both spectra.
Take $n=4$ for instance, where one finds the degeneracies 1-3-3-1-3-5 in ascending values of the CH spectrum.
In Fig. \ref{fig:su2_ed_dmrg}(c) one finds the identical sequence of degeneracies at $n=4$, in ascending order of the  entanglement spectrum.
The correspondence in the degeneracy hierarchy that exists between the corner spectrum and the entanglement spectrum for each $n$ is symbolically expressed,

\ba ({\rm ES})^{(n)} \propto ({\rm CS})^{(n)} . \ea
Later we will show that the correspondence is even stronger,  in that the exact positions of each level match between the two spectra. We summarily express this observation as

\ba ({\rm ES})^{(n)}  = \kappa ({\rm CS})^{(n)} , \ea
where $\kappa$ is some model-specific scaling factor. The correspondence holds not only for critical spin chains, but also for other well-known integrable models, for each $n$. It may be wondered how the comparison would fare if one used the Hamiltonian of length $n$, $H^{(n)}_{\rm SU(2)}  = \sum_{i=1}^{n-1} h_{\text{SU}(2),i}$, instead of the corner Hamiltonian of the same length. Such comparison is shown in the Appendix. As the reader can see, the degeneracy count for each level still shows excellent correspondence, but the level spacings do not agree well between the ES and the Hamiltonian spectra.

Next we examine the EE of the SU(2) chain.
The EE plot presented in Fig. \ref{fig:su2_ed_dmrg}(b) has been obtained several times in the past~\cite{affleck06,hastings09,xavier11,taddia11,kaul15} in various spin-1/2 chain systems with OBCs.
According to Refs. \onlinecite{calabrese04,calabrese10}, the EE in (1+1)-dimensional CFT with open boundaries
obeys the formula~\cite{kaul15}:

\begin{align}\label{eq:1+1cft}
  S_n &= S_n^{\rm CFT} + S_n^{\rm osc.} + c', \nn
  S_n^{\rm CFT} &= \frac{c_N}{6} \log \left[ \frac{2 L}{\pi} \sin \Big( \frac{\pi n}{L} \Big) \right], \nn
  S_n^{\rm osc.} &= \sum\limits_a F^a \!\!\left(\frac{n}{L} \right)  \frac{\cos ( 2 a \pi n /  N )}{
  \, \left| L\sin ( \pi n / L ) \right|^{\Delta_a} } ,
\end{align}
where $S_n$ is the von Neumann entropy of the subsystem of length $n$, $c_N = N-1$ and $\Delta_a$ are the central charge and scaling dimensions of the SU$(N)_1$ Wess-Zumino-Witten theory, $L$ is the system size, $c'$ is a non-universal constant, and finally $F^a \!(n/L)$ is the universal scaling function which can be adequately approximated as a constant~\cite{kaul15}. For $N=2,3$ there is only one scaling dimension coming into play $(a=1)$, but for $N=4$ there exist two distinct scaling dimensions $(a=1,2)$ to account for.  An excellent fit to the numerically obtained EE is possible by adjusting the constants $c'$ and $F^a$ as seen in Fig. \ref{fig:su2_ed_dmrg}(b).

It is natural to suspect that the origin of the oscillating EE is the oscillation of the entanglement spectrum itself.
As shown in Fig. \ref{fig:su2_ed_dmrg}(c), the ES settles into a period-2 structure when sufficiently removed from the edge.
For odd $n$ the lowest entanglement level is doubly degenerate and all other levels have the even degeneracy as well.
For even $n$ the lowest entanglement level is unique and other levels have the odd degeneracy. A qualitative understanding of such structures will be given in the next subsection.

At first sight it might seem that period-2 oscillations in the ES is compatible with the alternating term in the EE. A close inspection of our numerical data shows, on the other hand, that the alternating structure in the ES \emph{does not} diminish deep inside the chain, while the corresponding oscillation in the EE is known to decay algebraically as $n$ approaches the center of the chain~\cite{affleck06}. It seems puzzling, even contradictory,  that EE has decaying oscillations while the underlying ES does not.
It also appears that the persistence of the period-2 ES oscillation is counter to the notion that boundary effects should decay algebraically away from the edge in a gapless system. To obtain some understanding of the apparent mismatch we analyze the site-dependent WES defined earlier in Eq. (\ref{eq:WES}) for each location $n$ of the decomposition. The entanglement spectrum  is obtained by summing over all the WES's.
The WES for each cut position $n$ is plotted in Fig. \ref{fig:su2_ed_dmrg}(d) with corresponding ES levels in matching colors.
For $n$ large (deep in the bulk) and odd, only the two lowest entanglement levels, each doubly degenerate, contribute much to the entanglement entropy because they are the only ones with significant WES. For $n$ large and even, the lowest singlet and the next-lowest triplet levels contribute dominantly to the entanglement entropy, while all other levels have WES's very close to zero.
It is then entirely conceivable that two doublets at odd $n$ and singlet$+$triplet at even $n$ have their WES levels distributed in such a crafted manner that their contributions to the entanglement entropy become similar.  One can notice in Fig. \ref{fig:su2_ed_dmrg}(d) a mild growth in the overall WES levels with increasing $n$, in agreement with the growing EE towards the middle of the chain.
Overall, one can reconcile the asymptotically uniform behavior of the entanglement entropy with the persistent oscillation of the entanglement spectrum by a close inspection of the behavior of WES.

Nevertheless, the persistent even-odd alternation of the ES or WES appears to run counter to the intuitive notion of
decaying boundary effects that govern the behavior of physical observables such as the bond energy $\langle \bm \sigma_i \cdot \bm\sigma_{i+1}\rangle$~\cite{affleck06}. Additionally, we find it surprising that the entanglement entropy is dominated by only a handful of entanglement levels, and furthermore the inspection of small-$n$ (see WES for $n=2$ and $n=3$) WES structure is already adequate to account for the relevant WES for all following $n$'s (see WES for all $n \ge 4$).

In caveat, we point out that the site-by-site entanglement spectra and the corner Hamiltonian spectra shown in Fig.
\ref{fig:su2_ed_dmrg} and all other subsequent figures refers to the low-lying sectors only, as $n$ becomes sufficiently large and one is unable to display all the levels. It should also be emphasized, on the other hand, all the levels that are relevant for the calculation of the entanglement entropy are fully displayed in our figures irrespective of the value of $n$.

\subsection{Young tableaux consideration}

\begin{figure}[htbp]
  \centering
    \includegraphics[width=0.45\textwidth]{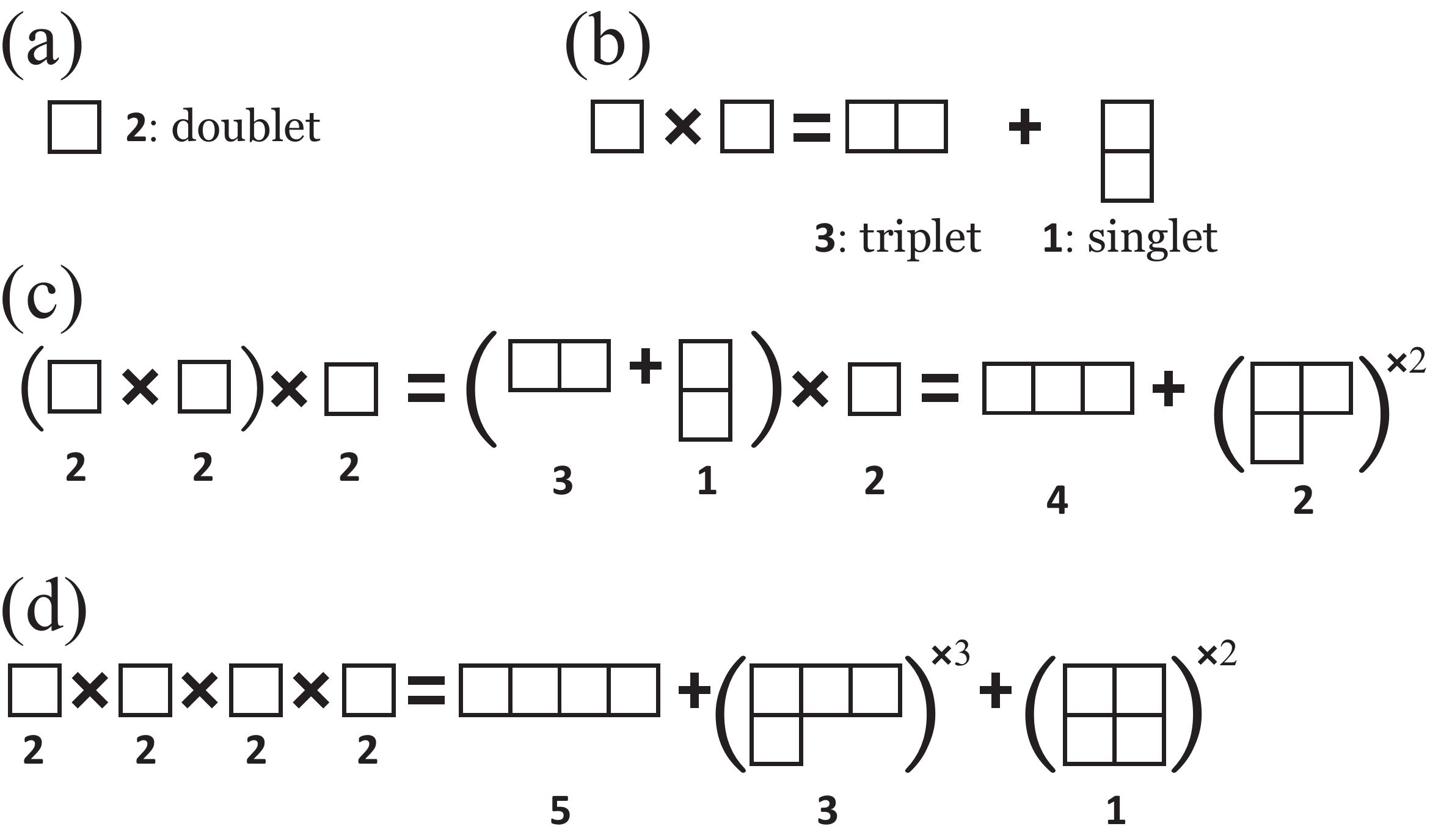}
   \caption{(a)-(d) Young tableaux for addition of spins in the left subsystem of $n=1,2,3,4$ cuts, respectively. The degeneracies in the ES correspond to dimensionalities of Young tableaux displayed as boldfaced numbers. Superscripts $^{\times 2}$ and $^{\times 3}$ indicate how many times a given representation appears.}
\label{fig:young_su2}
\end{figure}

We show that it is possible to obtain a qualitative understanding of the ES in the ground state of SU(2) critical spin chain using the Young tableaux consideration~\cite{greiter07,greiter08,thomale15}. For SU(2), a single tableau represents a spin doublet. When a cut is made between sites 1 and 2, namely $n=1$, there exists a single spin in the left subsystem whereas all the other spins belong to the right subsystem. We label a spin (or a doublet) by $\v 2$ as shown in Fig. \ref{fig:young_su2}(a).
To minimize the energy of the Heisenberg Hamiltonian, the spin of the total system should be a singlet when the chain length $L$ is even, as also dictated by the Marshall-Lieb-Mattis theorem~\cite{MLM}. Since $\v 2$ is in the left subsystem, the only way for the total system to achieve $\v 1$, a singlet, is to have another $\v 2$ in the right subsystem and make a singlet combination of the two $\bm 2$'s, $\bm 2 \otimes \bm 2 \rightarrow \bm 1$. In other words even though there exist $(N-1)$ number of spins in the right subsystem, it still acts effectively as a single spin-1/2 object, i.e. a $\v 2$. Following this idea one can write down the many-body ground state of the chain as

\begin{align}\label{eq:su2_singlet}
  |{\rm GS}_{\rm SU(2)} \rangle &= \frac{1}{\sqrt{2}} |\uparrow^{(l)} \rangle |\downarrow^{(r)} \rangle -
     \frac{1}{\sqrt{2}} |\downarrow^{(l)} \rangle |\uparrow^{(r)} \rangle \nn
     & = \frac{1}{\sqrt{2}} |\uparrow^{(l)} \rangle |\downarrow^{(r)} \rangle  +
     \frac{1}{\sqrt{2}} |\downarrow^{(l)} \rangle |- \uparrow^{(r)} \rangle  ,
\end{align}
where $|\uparrow^{(r)}\rangle$, $|\downarrow^{(r)}\rangle$ symbolically expresses the  composite state of the $(N-1)$ spins in the right subsystem with the total spin projection $\pm 1/2$, and $(l)$ refers to the left subsystem. From the second line, where we defined $|-\! \uparrow^{(r)} \rangle   \!=\! -  | \! \uparrow^{(r)} \rangle$, one can read off the two degenerate Schmidt eigenvalues  $\lambda_1=\lambda_2=1/\sqrt{2}$.

Moving on to the $n=2$ cut, now we have two spin-1/2's in the left subsystem forming $\v 1$ (singlet)
and $\v 3$ (triplet) as shown in Fig. \ref{fig:young_su2}(b).  Both $\v 1$ and $\v 3$ can form the total spin singlet with the composite spin on the right side either by $\v 1 \otimes \v 1 = \v 1$ or $\v 3 \otimes \v 3 = \v 5 \oplus \v 3 \oplus \v 1$.
The way to form $\v 1$ out of two $\v 3$'s is

\begin{align}\label{eq:3+3=1}
  |\v 1\rangle &= \frac{1}{\sqrt{3}} \left(  |1^{(l)}, -1^{(r)} \rangle - |0^{(l)}, 0^{(r)} \rangle + |-1^{(l)}, 1^{(r)} \rangle \right) .
\end{align}
The corresponding reduced density matrix for the left subsystem is easily worked out, $\rho^{(l)} = {\rm diag}(1/3,1/3,1/3)$, leading to the triple degeneracy in the Schmidt eigenvalue shown in Fig. \ref{fig:su2_ed_dmrg}(c) at $n=2$. The single ES level follows of course from $\v 1 \otimes \v 1 = \v 1$. The ground state can be written symbolically in the Schmidt form,

\ba\label{eq:young-su2_sch2}
  |{\rm GS}_{\rm SU(2)\!} \rangle  = \lambda^{\v 1}  |\v 1^{(l)} \rangle |{\v 1}^{(r)} \rangle
+ \sum_{\alpha=1}^3 \lambda^{\v 3}_\alpha |\v 3_\alpha^{(l)} \rangle |\v 3_{\alpha}^{(r)} \rangle .
\ea
On energetic grounds one expects the singlet part of the left wave function $|\v l^{(l)}\rangle$ to have more weight than the triplet part $|\v 3^{(l)}\rangle$ in the make-up of the overall ground state, i.e. $\lambda_1^{\v 1} > \lambda_1^{\v 3}=\lambda_2^{\v 3}=\lambda_3^{\v 3}$. After all, the triplet state $|\v 3^{(l)}\rangle$ is the excited state of the left subsystem and less likely to participate in the formation of the overall ground state. Inspection of the WES reveals, on the other hand, that $- (\lambda_\alpha^{\v 3})^2 \log (\lambda_\alpha^{\v 3})^2 \simeq - (\lambda^{\v 1})^2 \log(\lambda^{\v 1})^2 $. Since there are three states in the triplet, we find roughly three times as much contribution to the EE from the triplet Schmidt states as the singlet state.

Finally, consider the ES at $n=3$. In the left subsystem we have one quartet and two doublets as shown in the Young tableaux diagram, Fig. \ref{fig:young_su2}(c). The degeneracy of ES levels in Fig. \ref{fig:su2_ed_dmrg}(c) for $n=3$ coincides precisely with the dimensionality of each Young tableaux. Expressed as a Schmidt form at $n=3$, the ground state reads

\begin{align}\label{eq:young-su2_sch3}
  |{\rm GS}_{\rm SU(2)} \rangle &=
    \sum_{\alpha=1,2} \lambda_\alpha^{\v 2} |\v 2^{(l)}_\alpha \rangle |\v 2_{\alpha}^{(r)} \rangle \nn
   &+ \sum_{\alpha=1,2}  \lambda_\alpha^{\v 2'} |\v 2'^{(l)}_\alpha \rangle |\v 2'^{(r)}_\alpha \rangle \nn
   &+ \sum_{\alpha=1}^4 \lambda_\alpha^{\v 4} |\v 4^{(l)}_\alpha \rangle |\v 4^{(r)}_\alpha \rangle .
\end{align}
Each of the Schmidt-decomposed states in the first two lines is expected to be of the same singlet form as Eq.\;(\ref{eq:su2_singlet}).
Based on the energetic consideration, one expects $\lambda_\alpha^{\v 2} > \lambda_\alpha^{\v 2'} > \lambda^{\v 4}_\alpha$.
Inspection of the WES shows that $\lambda^{\v 4}_\alpha$ makes negligible contribution to the EE.

Continued in this manner, we can understand the multiplicities of entanglement levels as well as their ordering for a given $n$ from the  combination of energetic and group-theoretical arguments.

\section{SU(3) and SU(4) critical spin chains}
\label{sec:SU(3),SU(4)}

\begin{figure*}[htbp]
  \centering
    \includegraphics[width=0.90\textwidth]{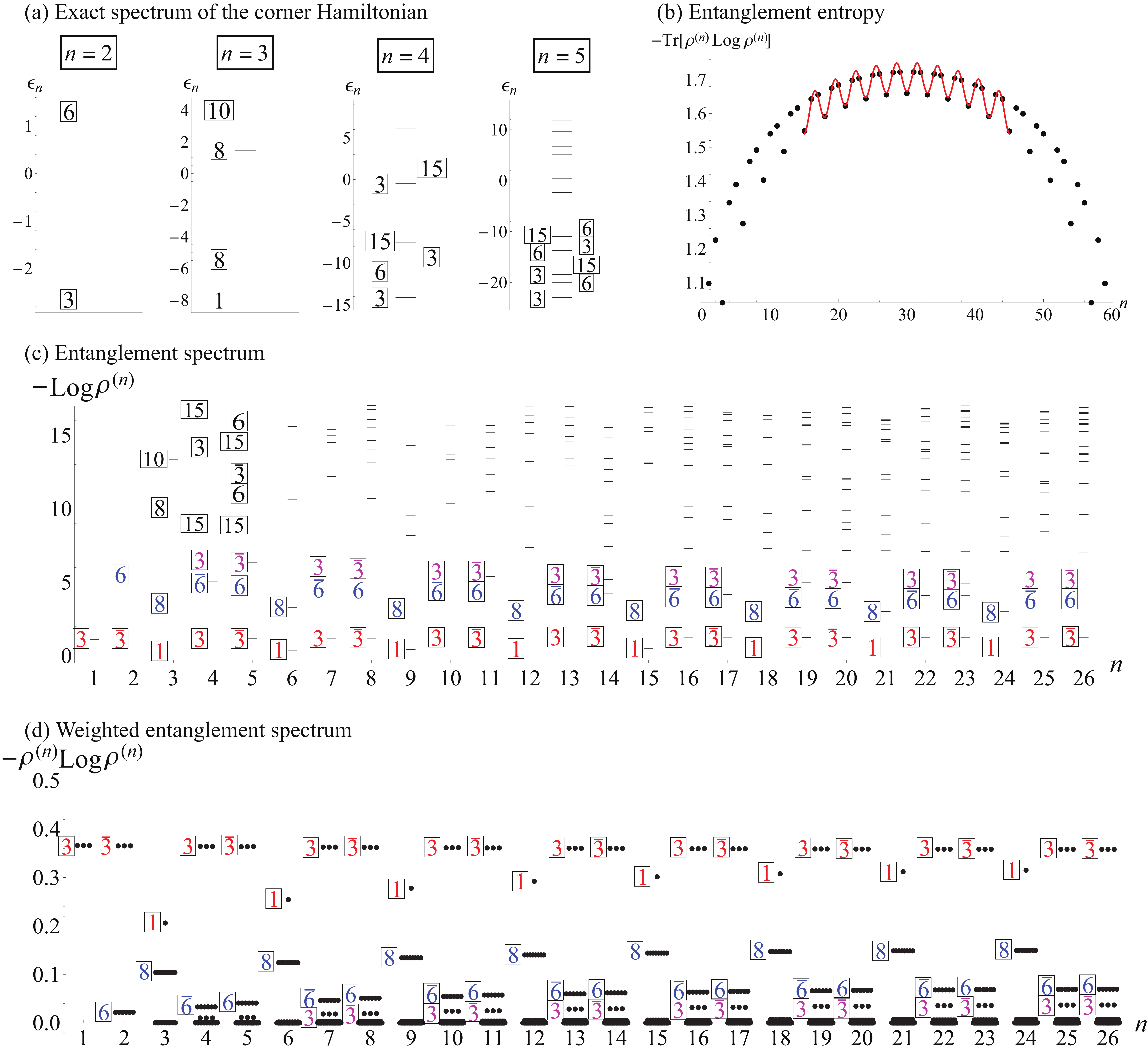}
   \caption{(color online) SU(3) critical spin chain.  (a) Spectrum of SU(3) CH of length $n$.
   (b) EE, (c) ES, and (d) weighted ES at the $n$-th cut of the ground state for a spin chain of length 60.
   The degeneracy of the levels in (a), (c), (d) is indicated inside the box next to the horizontal bar. The red solid fit in (b) is obtained from  Eq. (\ref{eq:1+1cft}) with $c_N=2$ and $\Delta_1=2/3$. The ES alternate over three sites in the bulk.
EE at mod$(n,3)=0$ is dominated by a singlet and an octet,
at mod$(n,3)=1$ by $\v 3, \bar{\v 6}, \v 3$, and at mod$(n,3)=2$ by $\bar{\v 3}, \v 6, \bar{\v 3}$.
Other entanglement levels have WES values that accumulate around zero as shown in (d). }
\label{fig:su3_ed_dmrg}
\end{figure*}

\begin{figure}[htbp]
  \centering
    \includegraphics[width=0.45\textwidth]{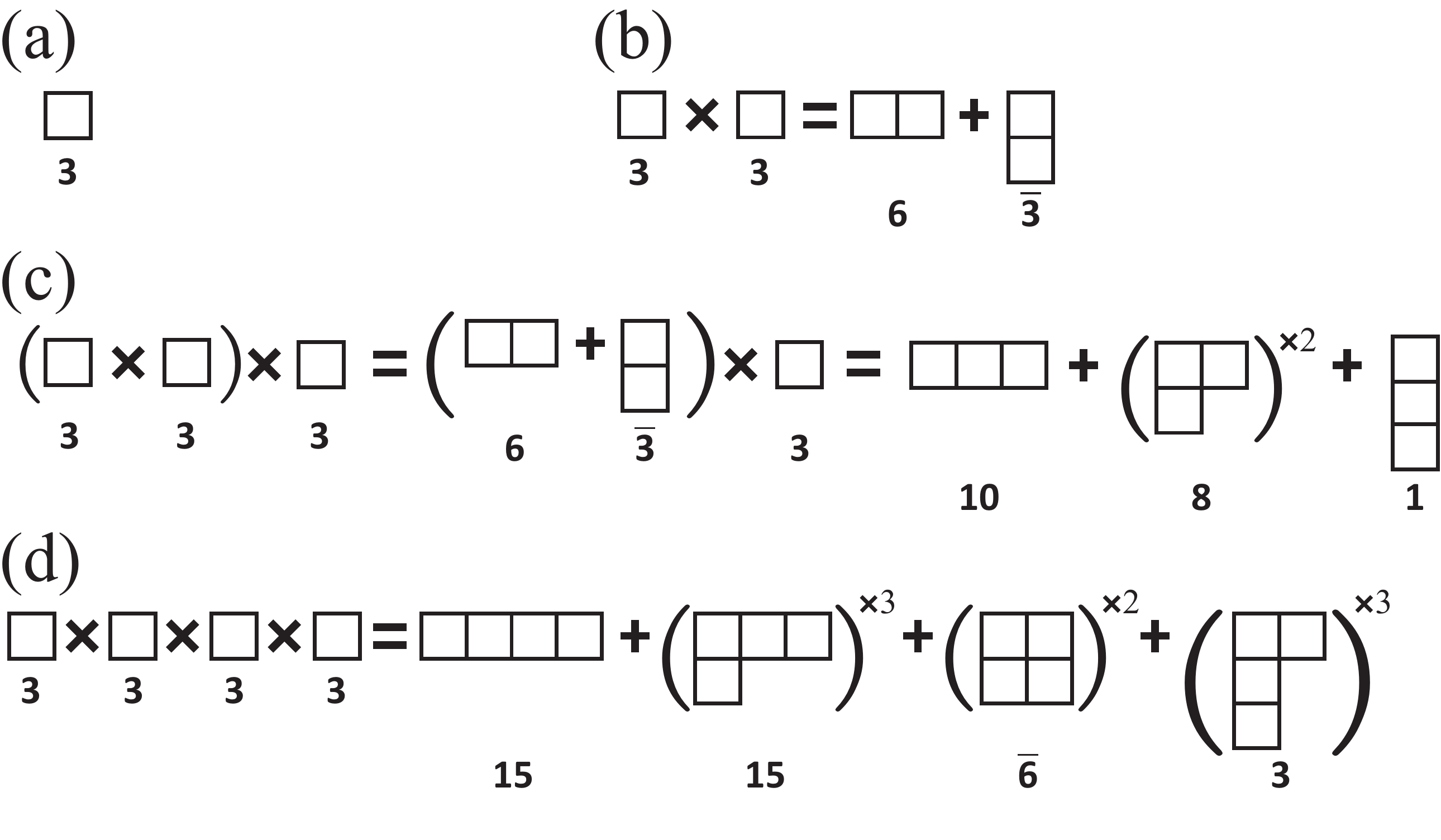}
   \caption{(a)-(d) Young tableaux for primitive objects of SU(3) corresponding to $n=1,2,3,4$ cuts, respectively. The degeneracies in ES correspond to dimensionalities of the Young tableaux. Conjugate representations are denoted by a bar. Superscripts $^{\times 2}$ and $^{\times 3}$ indicate how many times a given representation appears.}
\label{fig:young_su3}
\end{figure}

The SU(3) analogue of the SU(2) critical spin chain is the Uimin-Lai-Sutherland (ULS) model after the authors of three papers that contributed significantly to its Bethe ansatz
solution~\cite{uimin,lai,sutherland}. The Hamiltonian of the SU(3) critical spin chain of length $L$ is written as

\begin{align}
  &H^{(L)}_{\rm SU(3)} = \sum_{i=1}^{L-1} h_{\text{SU(3)},i} , \nn
  &h_{\text{SU(3)},i} = \bm \Lambda_i \cdot \bm \Lambda_{i+1},
\label{eq:su(3)-model}
\end{align}
where $\bm \Lambda = (\Lambda^1 , \cdots , \Lambda^8)$ is a collection of eight Gell-Mann matrices.
In this section, we perform DMRG calculations for SU(3) and SU(4) models on $L=60$ length spin chains.

In Fig. \ref{fig:su3_ed_dmrg}(a) we plot the spectrum of the SU(3) CH $K^{(n)}_{\rm SU(3)} = \sum_{i=1}^{n-1} i \, h_{\text{SU(3)},i}$. One can readily identify the perfect correspondence in the degeneracy of the spectrum of the CH with that of the ES shown in Fig.\;\ref{fig:su3_ed_dmrg}(c) for the same $n$.  The correspondence with the subsystem Hamiltonian $H^{(n)}_{\rm SU(3)} = \sum_{i=1}^{n-1} \, h_{\text{SU(3)},i}$ is worse (we do not show the calculation result separately).

In Fig. \ref{fig:su3_ed_dmrg}(b), the EE obtained from DMRG as a function of $n$ is plotted, with a good CFT fit using $c_N=2$ and scaling dimension $\Delta_1=2/3$ in Eq.\;(\ref{eq:1+1cft}). A period-3 oscillation is clearly observed.
The ES shown in Fig.\;\ref{fig:su3_ed_dmrg}(c) reveals the three-site alternation well into the bulk. Finally, Fig.\;\ref{fig:su3_ed_dmrg}(d) shows the WES as a function of $n$. As in the SU(2) case, only a small number of entanglement levels contribute significantly to the EE. The distinct patterns of the WES in the bulk still give rise to similar EE
after summation over the few significant levels. Period-3 oscillations of EE were reported in the spin-1 bilinear-biquadratic chain~\cite{solyom07} and SU(3) Hubbard model~\cite{solyom08}, but the periodic structure in the ES was overlooked in those studies.

The Young tableaux consideration provides an understanding of the degeneracies in the ES of the SU(3) chain. The ES of the SU(3) chain at $n=1,2,3,4$ cuts in Fig.\;\ref{fig:su3_ed_dmrg}(c) again correspond to dimensionalities of the SU(3) Young tableaux shown in Fig.\; \ref{fig:young_su3}. It must be borne in mind that in SU(3) both $\v 3$ (a single tableau) and its conjugate $\bar{\v 3}$ (two vertically placed tableaux) have the same dimensionality of 3. For sites satisfying mod$(n,3)=0$, the SU(3) singlet $\v 1$ and an octet $\v 8$ dominate the WES as seen in Fig.\;\ref{fig:su3_ed_dmrg}(d). Readers are referred to the diagram in Fig.\;\ref{fig:young_su3}(c) for the appearance of $\v 1$ and $\v 8$ when three blocks are pieced together. At the next sites where mod$(n,3)=1$, one can refer to the diagram in Fig.\;\ref{fig:young_su3}(d) to find $\v 3, \bar{\v 6}, \v 3$, among other representations. These three make up most of the EE contributions as one can see from the WES distribution in Fig.\;\ref{fig:su3_ed_dmrg}(d) at mod$(n,3)=1$. For mod$(n,3)=2$ one can easily generate the relevant Young tableaux from combining Figs.\;\ref{fig:young_su3}(c) with \ref{fig:young_su3}(b), which yields representations $\bar{\v 3}, \v 6, \bar{\v 3}$ among others. They dominate the WES distribution at mod$(n,3)=2$ sites with nearly the same WES values as the $\v 3, \bar{\v 6}, \v 3$ at mod$(n,3)=1$. In other words, the seemingly identical WES distributions at mod$(n,3)=1$ and mod$(n,3)=2$ are coming from representations that are conjugate with each other.

%
%

\begin{figure*}[htbp]
  \centering
    \includegraphics[width=0.90\textwidth]{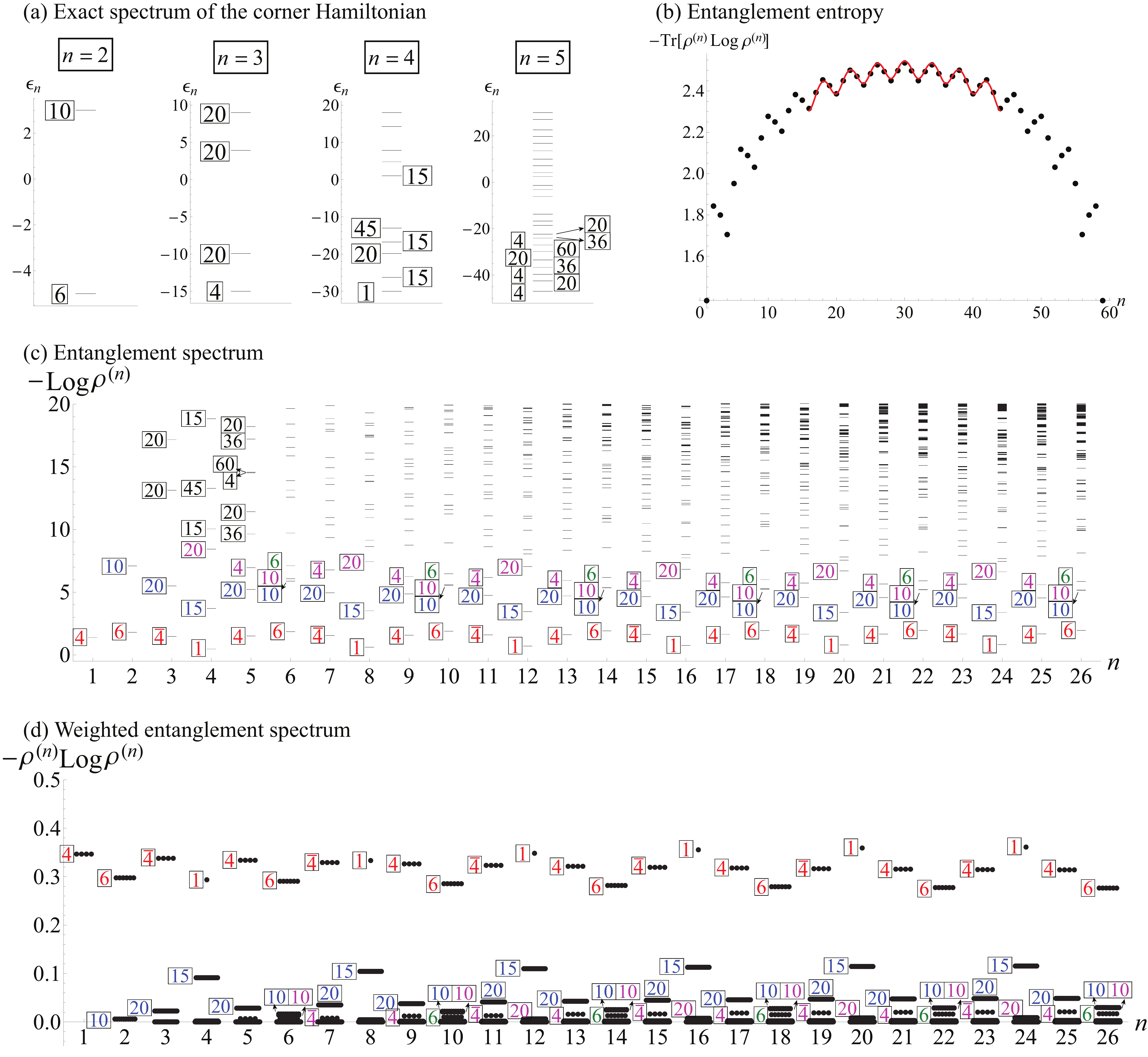}
   \caption{(color online) SU(4) critical spin chain. (a) Spectrum of SU(4) CH of length $n$.
   (b) EE, (c) ES, and (d) weighted ES at the $n$-th cut of the ground state for a spin chain of length 60.
   Degeneracy of the levels in (a), (c), (d) is indicated inside the box next to the horizontal bar.
   The ES alternate over four sites in the bulk.}
\label{fig:su4_ed_dmrg}
\end{figure*}

\begin{figure}[htbp]
  \centering
    \includegraphics[width=0.45\textwidth]{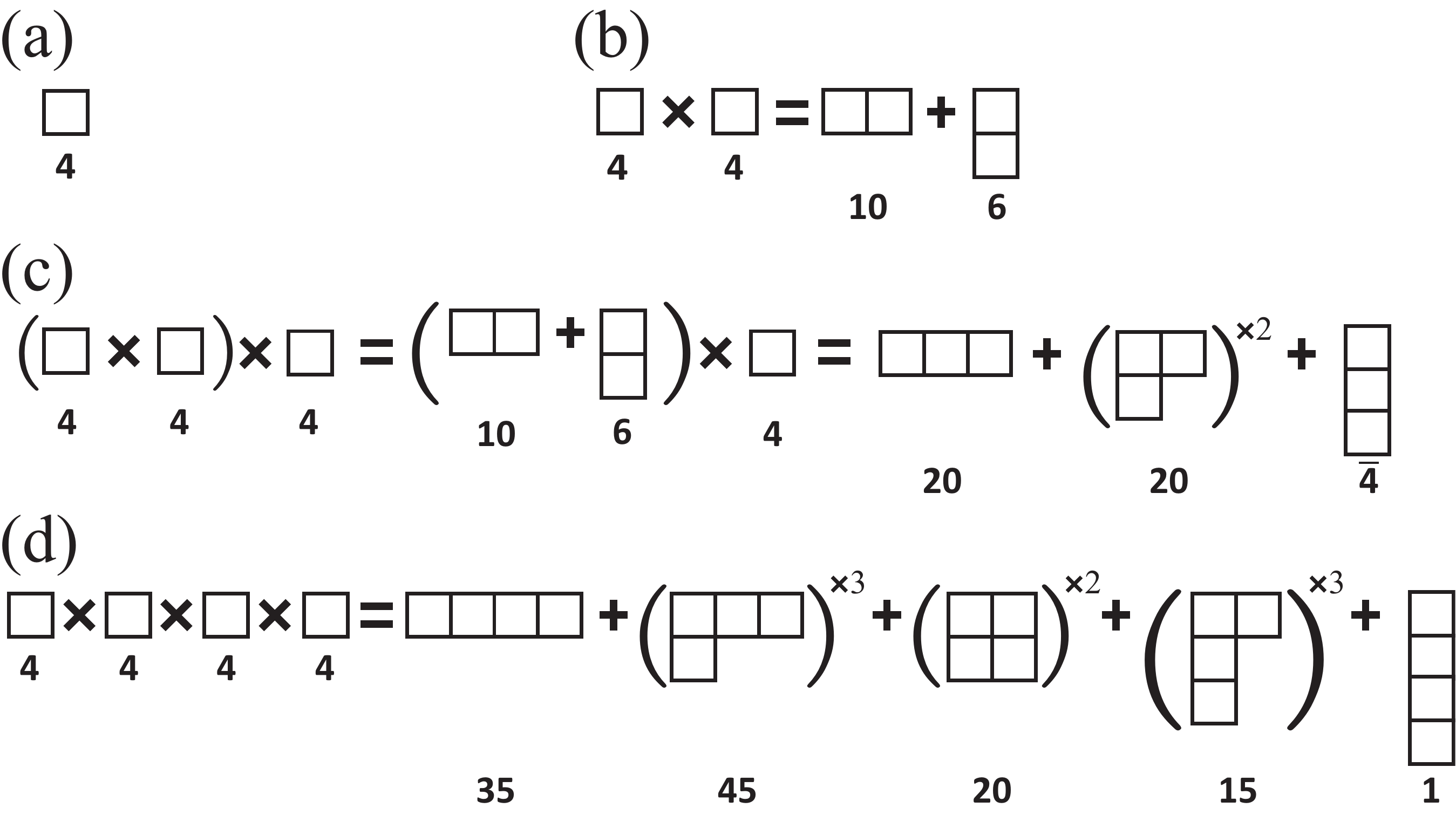}
   \caption{(a)-(d) Young tableaux for primitive objects of SU(4) corresponding to $n=1,2,3,4$ cuts, respectively. The degeneracies in ES correspond to dimensionalities of Young tableaux.}
\label{fig:young_su4}
\end{figure}

We next discuss the well-known Hamiltonian possessing the SU(4) symmetry, the Kugel-Khomskii model~\cite{KK},

\ba H_{\rm KK}^{(L)} =\sum_{i=1}^{L-1} (\bm \sigma_i \cdot \bm \sigma_{i+1}+1 )(\bm \tau_i \cdot \bm \tau_{i+1} +1) .\ea
In the original physical context, $\bm \sigma$ and $\bm \tau$ are both Pauli matrices referring to spin and orbital degrees of freedom, respectively.
Spin-orbital EE and ES of Kugel-Khomskii model with periodic boundary condition have been studied in Ref. \onlinecite{rex12}, but their main concern was to examine the boundary effect on quantum entanglement.
As noted in Ref. \onlinecite{KK-su(4)}, this model has an equivalent expression

\begin{align}
  &H^{(L)}_{\rm SU(4)} = \sum_{i=1}^{L-1} h_{\text{SU(4)},i} , \nn
  &h_{\text{SU(4)},i} = \bm \Gamma_i \cdot \bm \Gamma_{i+1}
\end{align}
in terms of the fifteen gamma matrices $\bm\Gamma = (\Gamma^1 , \cdots, \Gamma^{15})$ that generate the SU(4) algebra.

Shown in Fig.\;\ref{fig:su4_ed_dmrg}(a) are the spectrum of the SU(4) CH $K^{(n)}_{\rm SU(4)} = \sum_{i=1}^{n-1} i \, h_{\text{SU(4)},i}$ consisting of $n$ sites with OBC.
The ground state of $H_{\rm SU(4)}$ with 60 sites is obtained by DMRG, and its EE, ES, and WES are plotted in Fig.\;\ref{fig:su4_ed_dmrg}(b), (c), and (d), respectively.
Again the degeneracy of the spectrum of the CH and the ES at the same $n$ match well.
In Fig.\;\ref{fig:su4_ed_dmrg}(b), the EE as a function of $n$ along with (1+1)-dimensional CFT fit is presented with $c_N=3$ and two scaling dimensions $\Delta_1=3/4$ and $\Delta_2=1$.
The period-4 oscillation in the EE originates from the ES structure shown in Fig.\;\ref{fig:su4_ed_dmrg}(c).
Only a handful of the ES contribute significantly to the EE as one can see from the inspection of the WES in Fig.\;\ref{fig:su4_ed_dmrg}(d). In Ref.~\onlinecite{solyom08}, the one-dimensional SU(4) Hubbard model has been studied showing a similar EE oscillation, but the behavior of the ES was not investigated. Once again, we find that the Young tableaux analysis as shown in Fig.\;\ref{fig:young_su4} accounts for degeneracies in the ES and the period-4 oscillation.

\section{Corner Hamiltonians and entanglement spectra of integrable models}
\label{sec:CH-ES}

In the previous two sections, we have examined the entanglement structures of the SU($N$) critical spin chains and compared them with the spectra of the associated corner Hamiltonians.
We have noted the remarkable one-to-one correspondence in the degeneracy structures of the CS and the ES for subsystems of the same size $n$. In this section, we elevate this observation by showing that even the exact level positions match between the two spectra, and that the match occurs for many other integrable models, whether gapped or not.

Although Eq.~(\ref{eq:CH-ES_connection}) was expected to hold only in semi-infinite systems~\cite{peschel98}, our numerical methods can directly compare the spectrum of CH of arbitrary finite length $n$ with ES at the cut position $n$ of the ground state of physical Hamiltonian. Quite remarkably, we find a near perfect correspondence in the two spectra for arbitrary $n$. The length of the chains in the DMRG simulation is 60 throughout this section.

\begin{figure}[htbp]
  \centering
    \includegraphics[width=0.45\textwidth]{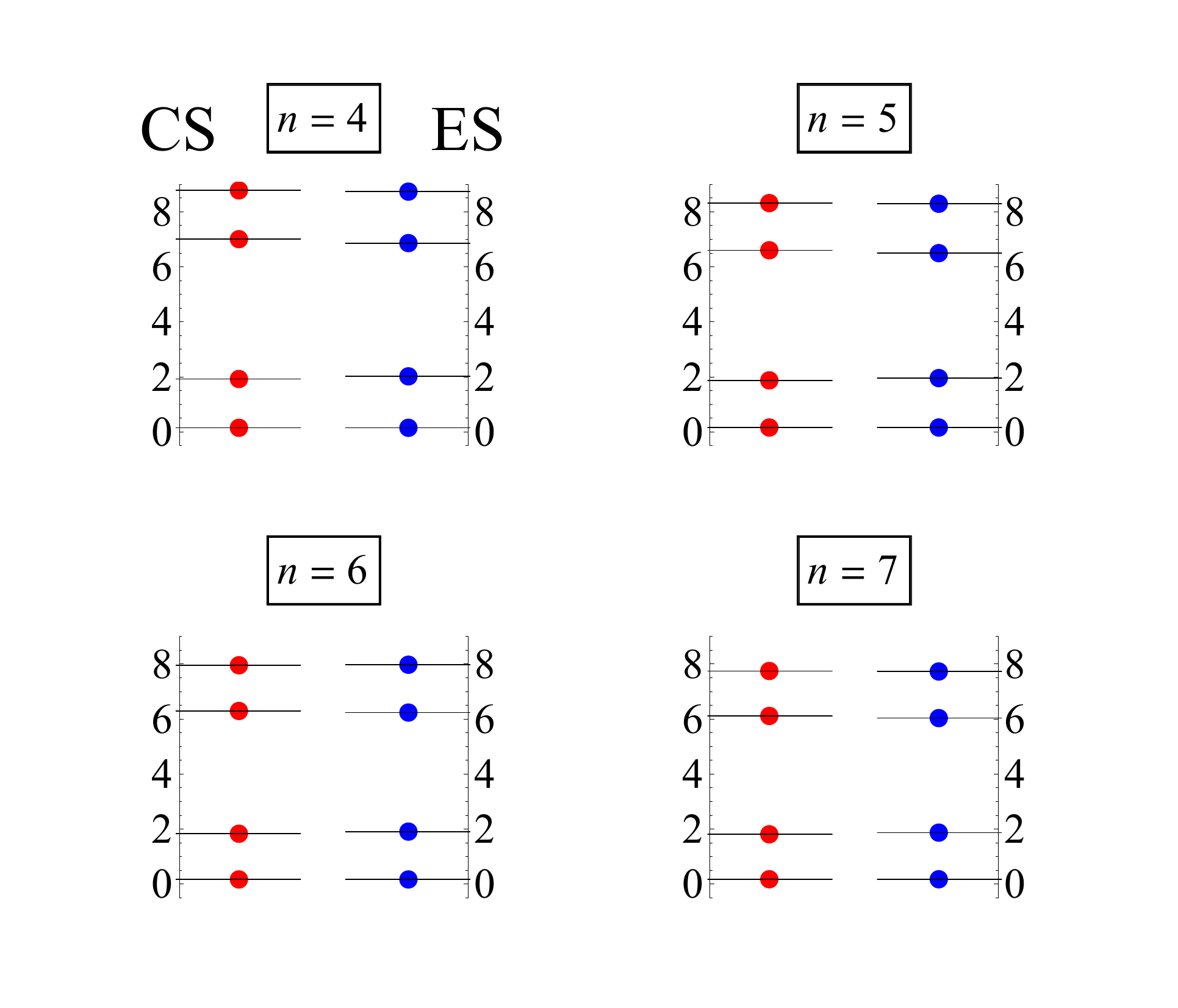}
   \caption{(color online) CS (left in each figure) and ES of the ground state of transverse Ising model (right in each figure) at the critical point $\lambda=1$. Spectrum of the CH is appropriately scaled to match with the ES, with the lowest eigenvalue of the CH being equal to that of the ES. The degeneracy of each level is indicated by the number of dots. The level structures in CS and ES are almost identical to each other.}
\label{fig:CH-ES_TIM}
\end{figure}

First, consider the transverse Ising model

\begin{align}
  H_{\rm TIM}= - \sum\limits_{i=1}^{L} \sigma_i^x - \lambda \sum\limits_{i=1}^{L-1} \sigma_i^z \sigma_{i+1}^z.
  \label{eq:TIM-H}
\end{align}
The system is gapped, except for the critical points $\lambda=\pm 1$.
The corresponding CH reads~\cite{davies88}

\begin{align}
  K_{\rm TIM}= - \sum\limits_{i=1}^{n} (2i-1)\sigma_i^x - \lambda \sum\limits_{i=1}^{n-1} 2i \sigma_i^z \sigma_{i+1}^z,
  \label{eq:TIM-K},
\end{align}
whose spectrum has been obtained analytically using the orthogonal polynomial method~\cite{peschel89,davies90,peschel90}. A more quantitative analysis was carried out numerically for $\lambda=0.8$ and $\lambda=1.25$ in Ref.~\onlinecite{peschel98}.
The RDM of the TIM for the arbitrary bipartition is given by the formula~\cite{peschel01}

\begin{align}
  \rho^{(n)} \propto e^{-\sum\limits_{l=1}^n \epsilon_l f_l^\dag f_l },
\end{align}
where $f_l$ is a fermionic operator.
Here $\epsilon_l$ which gives rise to the ES is generally obtained from numerical studies, except in the thermodynamic limit away from criticality where the asymptotic form of the solution can be found~\cite{peschel98}.
The ES for $\lambda=1$ TIM has been studied numerically in Ref.~\onlinecite{peschel01} with a restriction $n=L/2$.
We have solved, on the other hand,  Eqs.~(\ref{eq:TIM-H}) and (\ref{eq:TIM-K}) by DMRG and exact diagonalization, respectively, to obtain the ES of the ground state and the spectrum of the CH for arbitrary $n$.
Figure \ref{fig:CH-ES_TIM} shows side-by-side comparisons of the CS and the ES of the TIM, at criticality $\lambda=1$, for the same $n \in \{ 4,5,6,7 \}$.
The degeneracy of the corner (entanglement) spectrum is denoted by the number of red (blue) dots on each level throughout the section.
In addition to having the same degeneracy structures, the level spacings in both spectra perfectly match after an overall scaling of one of them. The observation of identical level structures in CS and ES is surprising in the sense that the cut position $n \in \{ 4,5,6,7 \}$ plotted in Fig.~\ref{fig:CH-ES_TIM} is near the end of the chain where, normally, the boundary effect has to be taken account and lies outside the expected realm of the ES=CS correspondence.

\begin{figure*}[htbp]
  \centering
    \includegraphics[width=0.85\textwidth]{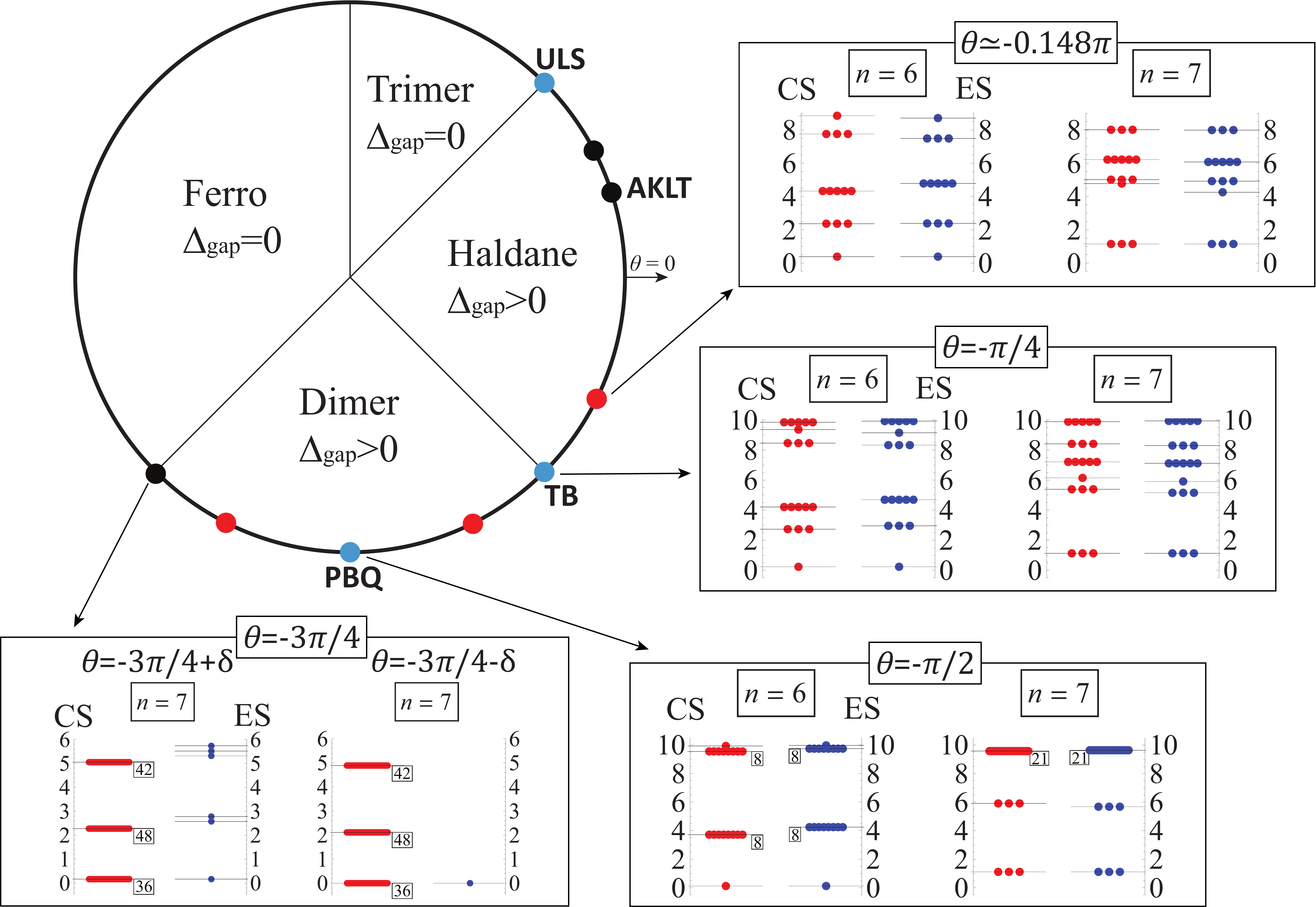}
   \caption{(color online) Phase diagram of spin-1 bilinear-biquadratic model and comparison of CS and ES at various $\theta$ values. Degeneracy of each level is indicated by the number of dots or inside the box next to the horizontal bar. The blue dots on the phase diagram indicate three integrable points, $\theta \in \{ -\pi/2, -\pi/4, \pi/4 \}$, where the level structures of the CS and the ES show one-to-one correspondence. The red dots on the phase diagram indicate non-integrable points at several $\theta$ values where the CS and the ES match reasonably well. The correspondence at non-integrable points is an open question. The black dots on the phase diagram are the points where the CS and the ES show completely different level structures; this behavior is currently not understood.
}
\label{fig:CH-ES_blbq}
\end{figure*}

Next we turn our attention to spin-1 bilinear-biquadratic model,

\begin{align}
  &~~~~~~~~~~~~~~ H^{(L)}_{\rm BLBQ} = \sum_{i=1}^{L-1} h_{\text{BLBQ},i} , \nn
  &h_{\text{BLBQ},i} = \cos \theta ( \v S_i \cdot \v S_{i+1} ) + \sin \theta (\v S_i \cdot \v S_{i+1})^2,
\label{eq:spin-1_blbq}
\end{align}
where three integrable points have been found at $\theta=\pi/4$ (ULS point~\cite{uimin,lai,sutherland}), $\theta=-\pi/4$ (Takhtajan-Babujian (TB) point~\cite{takhtajan,babujian}), and $\theta=-\pi/2$ (pure biquadratic (PBQ) point~\cite{parkinson,klumper89,barber89}).
Of course $\theta= \pm 3\pi/4$ are also integrable because they give minuses of TB and ULS Hamiltonians, respectively. Since their ground states are ferromagnetic and trivial, however, we exclude them from consideration. It is also the case the ground states are not unique for $\theta=\pm 3\pi/4$, whereas we are focused on models with unique ground states.

The corresponding corner Hamiltonian is defined as

\begin{align}
   K^{(n)}_{\rm BLBQ} = \sum_{i=1}^{n-1} i \, h_{\text{BLBQ},i}.
\label{eq:spin-1_CH}
\end{align}
Among the three integrable points, two of them at $\theta=\pm \pi /4$ (ULS and TB points) are gapless and critical, whereas the third one at $\theta=-\pi /2$ (PBQ point) is gapped.

Shown in Fig.~\ref{fig:CH-ES_blbq} is the phase diagram of spin-1 bilinear-biquadratic model, together with comparisons of the exact spectrum of Eq.~(\ref{eq:spin-1_CH}) and ES of the ground state of Eq.~(\ref{eq:spin-1_blbq}) with $L=60$ at various $\theta$ values from $-3 \pi /4$ to $\pi / 4$. The well-established phases of the BLBQ Hamiltonian are denoted inside the phase diagram~\cite{lauchli07}. The level structures of CS and ES at all integrable points, denoted by blue dots on the phase diagram, show excellent one-to-one correspondence. In particular, $\theta=\pi/4$ point is the ULS Hamiltonian of SU(3) symmetry already examined in the previous section.

In addition to the three integrable points, we have examined various other $\theta$ values at red-dotted points in Fig.~\ref{fig:CH-ES_blbq}. Surprisingly at $\theta \simeq -0.648\pi, -0.352\pi, -0.148\pi$ denoted by red dots on the phase diagram, the correspondence still holds even though those points are away from integrability.
The reason for such correspondence could be their proximity to integrable points. On the other hand, we found no such correspondence at $\theta \simeq 0.148 \pi$, represented as a black dot close to the ULS point. At the AKLT point $\tan \theta = 1/3$, the ES is a single level with double degeneracy whereas the corner Hamiltonian spectrum has many levels. The first-order phase transition point at $\theta=-3\pi/4$ also lacks the ES-CS correspondence, as shown in Fig. \ref{fig:CH-ES_blbq}.
\footnote{The degeneracies, 36, 48, 42, shown in the inset can be computed using the hook rule.}

\begin{figure}[htbp]
  \centering
    \includegraphics[width=0.48\textwidth]{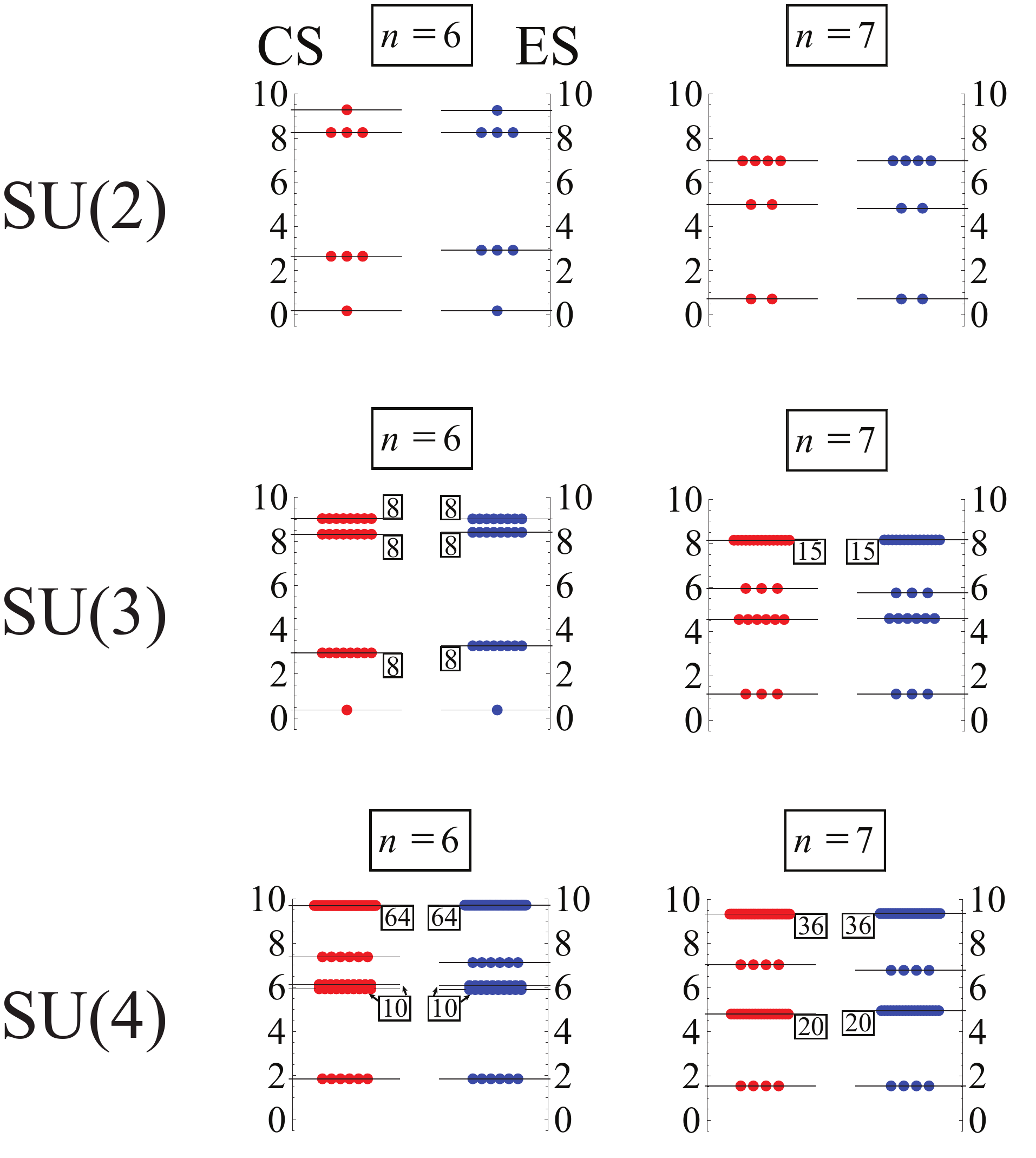}
   \caption{(color online) CS (left in each figure) of SU(2), SU(3), and SU(4) corner Hamiltonians for $n=6,7$ and ES (right in each figure) at $n=6,7$ of corresponding ground states. Degeneracy in each level is indicated by the number of dots or inside the box next to the horizontal bar. In addition to the same degeneracies of each spectrum, the level spacings become almost the same in all figures. }
\label{fig:CH-ES_suN}
\end{figure}

Finally, we return to the critical SU($N$) spin chains and compare the ES (already shown in the previous sections) with the corner spectrum of the Hamiltonians

\ba K^{(n)}_{\rm SU(N)} = \sum_{i=1}^{n-1} i \, h_{\text{SU(N)},i}  . \ea
%
As one can see in Fig.\;\ref{fig:CH-ES_suN}, the ES-CS correspondence is nearly perfect for all SU($N$) chains at arbitrary cut $n$ (only $n=6,7$ are shown for clarity) we examined.


\section{summary and discussion}
\label{sec:summary}

We have presented a thorough examination of the site-by-site entanglement structure of critical SU($N$) spin chains.
Specifically, we discuss the connection between the period-$N$ oscillations in the entanglement entropy of the SU($N$) critical spin chains with the site dependence of the entanglement spectrum. One of the most surprising findings of our DMRG investigation is the persistent oscillation of the ES well into the bulk. We introduce a new quantity called the weighted entanglement spectrum to facilitate the understanding of the ES behavior.

Going beyond the critical SU$(N)$ chain and covering other integrable models, we find robust evidence for the correspondence between the site-by-site ES and the corner Hamiltonian spectra of subsystems of the same size. The correspondence is found to hold also for a number of integrable points in the bilinear-biquadratic family of spin-1 chains with well-known unique ground states. The correspondence of the entanglement spectrum with those of the Hamiltonian, not the corner Hamiltonian, is generally worse in that while the degeneracies of the spectra match, the level spacing structures do not match. The period-$N$ oscillation of the ES we found for the critical SU$(N)$ model could be a special case of the general Hamiltonian with continuous spin-rotation symmetry.
It has been indirectly shown that the oscillation of the ES is a consequence of continuous symmetry~\cite{xavier12}.
We provided a simple counting argument based on the Young tableaux consideration that explains the origin of the period-$N$ behavior in SU($N$)-symmetric systems.

We conclude with remarks on open issues. First, existing ideas on the ES$=$CS correspondence rest on the technique of boundary CFT~\cite{lauchli13,cho16,cardy16}, which is a continuum theory, and are difficult to apply to arbitrary $n$ as the bipartition position.
A proof directly based on the lattice model calculation will be more in line with our observation, but presumably far more cumbersome than a continuum approach. In addition, our observation for the BLBQ family of Hamiltonians showed that the correspondence works even away from points of integrability. Hence the integrability of the model is not, strictly speaking, a necessary condition for the correspondence to be observed. How to mathematically formulate the condition for ES$=$CS correspondence is an interesting challenge from the point of view of both field theory and lattice models. Following the claim of Ref. \onlinecite{cho16}, it may be that lattice Hamiltonians for which the ES$=$CS correspondence is expected to have the low-energy description given by (1+1) gapped Lorentz invariant theories. How to tie such field theory analysis with observations made for lattice models is, however, beyond the scope of the present study.

Second, as we have seen, the correspondence $H^{(n)}_{\rm ent.} \propto K^{(n)}$ remarkably holds for all $n$ sufficiently smaller than the half-chain length $L/2$ though the derivation of the relation relies on a somewhat heuristic argument.
The result in the exactly solvable XX model suggests that the precise form of the entanglement Hamiltonian is rather $H_{\rm ent}^{(n)} = \sum_{i=1}^{n-1} i (2n-i+1)/(2n) \times h_i$~\cite{peschel04,peschel09}. CFT in special cases also predicts the parabolic variation of the energy density~\cite{cardy16}. Due to the inability to carry out exact diagonalization of the corner Hamiltonian for large values of $n$, we must leave open the question whether the prefactor $i (2n-i+1)/(2n)$, not $i$, is a more precise characterization of the entanglement Hamiltonian for all cut positions $n$.

Third, the Young tableaux argument we developed gives a heuristic foundation for understanding the ES oscillations in models with SU($N$) symmetry. We note that the same argument applies equally to spatially inhomogeneous chains as long as the SU($N$) symmetry remains intact. Whether a similar type of argument can be worked out
for more sophisticated symmetries such as the quantum group (e.g. $U_q (sl_2)$ to which the $S=1/2$ XXZ spin chain with boundary
magnetic fields belongs) is an issue we find interesting.

Last but not least, with rapid advances in the cold atom technology, perhaps it would be conceivable that the oscillations in the ES will find experimental manifestations in cold atom systems with SU($N$) symmetry~\cite{bloch12,totsuka16}. It will be particularly interesting to speculate on the experimental consequence of the {\it persistence of the ES oscillation} deep inside the SU($N$) chain.

\acknowledgments{J.H.H. would like to acknowledge enlightening inputs from Gil-Young Cho, Dung-Hai Lee, and Masaki Oshikawa. PK is grateful to Tomotoshi Nishino and Kouichi Okunishi for insightful comments. H.K. was supported in part by JSPS Grants-in-Aid for Scientific Research No.15K17719, No. JP16H00985, and 25400407. He acknowledges insightful discussions with John Cardy. N.T. acknowledges discussions with
Takahiro Morimoto and Sandip Trivedi, funding from NSF-DMR1309461 and partial support by a grant from the Simons Foundation (\#343227).
}

\appendix*

\section{Comparison of the Entanglement Spectrum with the Hamiltonian spectrum}

\begin{figure*}[htbp]
  \centering
    \includegraphics[width=0.95\textwidth]{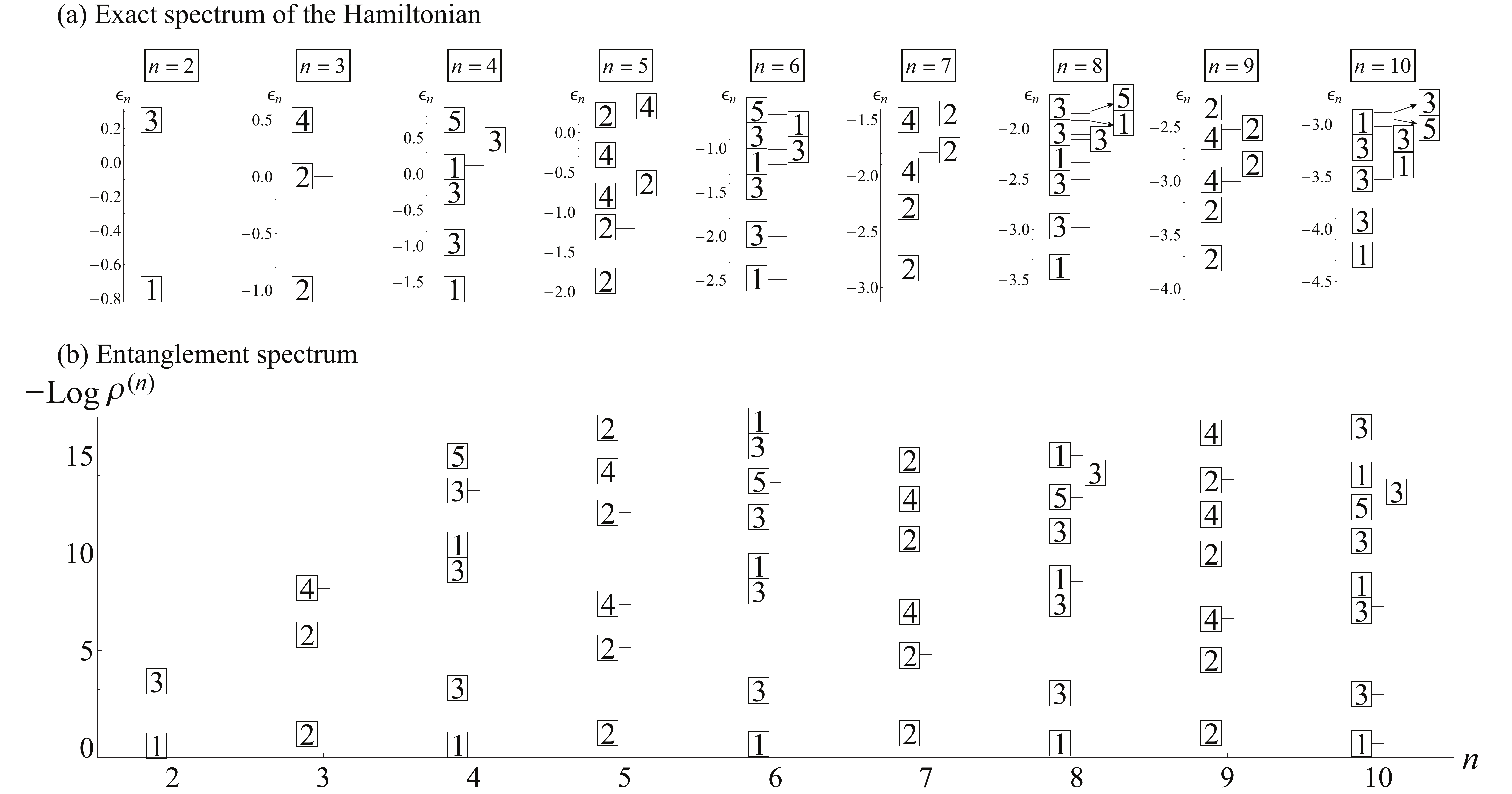}
   \caption{(color online) SU(2) critical spin chain. (a) Low energy spectrum of the SU(2) spin chain of length $n$.
   (b) The ES at the $n$-th cut of the ground state for a spin chain of length 60.
   Degeneracy of the levels is indicated inside the box next to the horizontal bar.}
\label{fig:PS-ES_su2}
\end{figure*}

One of the main messages of this work is that the corner spectrum CS and the entanglement spectrum ES match very well at any cut position $n$.
Here we further compare the ES with the energy spectrum of the physical Hamiltonian, not the corner Hamiltonian.

The energy eigenvalues of the SU(2) spin chains of length $L=n$, Eq.~(\ref{eq:SU(2)-model}), are obtained by exact diagonalization and shown in Fig.\;\ref{fig:PS-ES_su2}(a).
Up to $n=4$ the entire energy eigenvalues are shown but for $n \ge 5$ only a few low energy values are displayed for simplicity.
The ES obtained from the ground state of a spin chain of the same length $L=60$ are plotted up to cut position $n=10$ in Fig.\;\ref{fig:PS-ES_su2}(b).
Due to the global SU(2) symmetry which governs not only the Schmidt eigenstates but also those of the Hamiltonian, the degeneracy numbers are in good correspondence between the energy spectrum and the ES.
Unlike the CS, however, the spacings in the energy eigenvalues do not agree with those of the ES, and even the sequences of degeneracy levels are sometimes different.
For example, if we compare the energy spectrum and the ES at $n=6$, we see that upon
counting from the lowest levels, the degeneracy numbers are 1-3-3-1-3-\textbf{3-1-5} for the energy spectrum, but 1-3-3-1-3-\textbf{5-3-1} for the ES.

The poor correspondence between the energy and entanglement spectra is also observed in SU($N$) ($N=3,4$) chains, but not shown separately in this paper.

\bibliography{reference.bib}

\end{document}